\documentclass[11pt]{article}
\newcommand{\D}{\nabla}
\newcommand{\Fc}{\mathcal{F}}

\usepackage{amssymb,amsmath}

\begin{document}
\title{Cosmological Solutions in Nonlocal Models}

\author{Alexey~S.~Koshelev$^{1}$\footnote{Postdoctoral researcher of
FWO-Vlaanderen, E-mail:
alexey.koshelev@vub.ac.be} \ and \ Sergey~Yu.~Vernov$^{2}$\footnote{E-mail: svernov@theory.sinp.msu.ru} \vspace*{3mm} \\
{\small $^1$Theoretische Natuurkunde, Vrije Universiteit Brussel and The
International Solvay Institutes,}\\{\small  Pleinlaan 2, B-1050 Brussels,
Belgium}\\
{\small $^3$Skobeltsyn Institute of Nuclear Physics, Lomonosov Moscow State University,}\\
{\small Leninskie Gory 1, 119991, Moscow, Russia}}
\date{ }
\maketitle

\begin{abstract}
A non-local modified gravity model with an analytic function of the d'Alembert
operator that has been  proposed as a possible way of resolving the
singularities problems in cosmology is considered. We show that the anzats that
is usually used to obtain exact solutions in this model
provides a connection with $f(R)$ gravity models.
 \end{abstract}

\section{Introduction}

General relativity (GR) being a very efficient and simple theory of gravity
featuring the second order equations of motion suffers from certain weaknesses.
Among them the initial singularity problem which arises in the framework of the
inflationary paradigm. This is the reflection of UV incompleteness of GR.
One possibility to improve the ultraviolet behavior and even to get
a renormalizable theory of quantum gravity is
to add higher-derivative terms to the Einstein--Hilbert action. As one
of the
first papers we can mention~\cite{Stelle} where curvature squared corrections
were considered.
Unfortunately, this model (and models with more than two but finite number of
 derivatives in equations of motion in general) has ghosts. An intriguing
possibility to
overcome this problem is to consider a non-local gravity with infinitely many derivatives.

The theoretical motivation behind an introduction of infinite
derivative non-local corrections into a local theory is the string field theory
(SFT)~\cite{sft_review}. Such corrections naturally arise in the SFT
and usually consist of exponential functions of the d'Alembertian operator
acting on fields. The
majority of non-local cosmological models motivated by such
structures explicitly include an analytic or meromorphic
function of the d'Alembertian
operator~\cite{NLG,NLScF,BMS,BKM,BGKM,BKMV,KVSQS2011}.

Both GR and the modified gravity models yield
a non-integrable system of equations of motion with only particular analytic
solutions known. Having an analytic solution however is crucial in
considering perturbations which are in turn the cornerstone of any cosmological
model. Needless to say that finding analytic solutions in
non-local nonlinear equations is an extremely hard task. Some
studies of nonlocal modifications of GR resulted in analytic solutions can be
found in~\cite{BMS,BKM,KVSQS2011}.

The key to finding solutions in the non-local gravity models of interest is to
employ an ansatz which relates finite powers of the d'Alembertian operator
acting on the scalar curvature. The anzats itself reduces the initial
non-local model to an effective local model with more than two but finite number of derivatives. This does not mean that ghosts should
appear as any anzats is a relation in the background while perturbations
enjoy all the new properties of the full non-local structure.
Provided there is a background configuration satisfying the anzats one simplifies the problem of
solving the equations of motion considerably.

\section{Action and Equations of Motion for String-inspired Non-local Gravity}
\label{sec:background}

The nonlocally modified gravity proposed in~\cite{BMS} is described by the following action:
\begin{equation}
 S=\int d^4x\sqrt{-g}\left[\frac
 {M_P^2}{2}R+\frac{1}{2}R\Fc\left(\frac{\Box}{M_{\ast}^2}\right)R-\Lambda+ \mathcal{L}_\mathrm{M}\right],
 \label{nlg_action}
\end{equation}
where  $M_P$ is the Planck mass,
$\Lambda$  is the cosmological constant, $M_{\ast}$ is the mass scale at which the higher derivative
terms in the action become important, $\mathcal{L}_\mathrm{M}$ is the matter
Lagrangian. We use the convention where the metric $g$ has the signature $(-,+,+,+)$.
 $M_{\ast}$ is the mass scale at which the higher derivative
terms in the action become important.
An analytic function
$\Fc\left(\Box/M_{\ast}^2\right)=\sum\limits_{n\geqslant0}f_n\Box^n$ is an ingredient
inspired by the SFT. The operator $\Box$ is the covariant
d'Alembertian. In the case of an infinite series we have a non-local
action.

Introducing dimensionless coordinates
$\bar{x}_\mu=M_{\ast} x_\mu$ and $\bar{M}_P=M_P/M_{\ast}$ we get $\Fc(\Box/M_{\ast}^2)=\Fc(\bar{\Box})$, where $\bar{\Box}$ is
the  d'Alembertian in terms of dimensionless coordinates. We shall
use dimensionless coordinates only (omitting the bars).

A straightforward variation of action (\ref{nlg_action}) yields the
following system:
\begin{equation}
\begin{split}
\frac12[M_P^2+2\Fc(\Box)R]&\left(2R^\mu_\nu-\delta^\mu_\nu R\right)
=\frac{1}{2}\sum_{n=1}
^\infty
f_n\sum_{l=0}^{n-1}\Bigl[g^{\mu\rho}\partial_\rho\Box^l  R  \partial_\nu\Box^{n-l-1}  R
+{}\\
+& g^{\mu\rho}\partial_\nu\Box^l  R  \partial_\rho\Box^{n-l-1}  R  -\delta^\mu_{\nu}\left(g^{\rho\sigma}
\partial_\rho\Box^l  R  \partial_\sigma\Box^{n-l-1}  R  +\Box^l  R  \Box^{n-l}  R
\right)\Bigr]+{}\\+&2(g^{\mu\rho}\D_\rho\partial_\nu-\delta^\mu_{\nu}
\Box)\Fc(\Box) R-\frac{1}{2}
 R \Fc(\Box) R\delta^\mu_{\nu}-\Lambda \delta^\mu_{\nu}+{T}^\mu_\nu\, ,
\end{split}
\label{eqEinsteinRonlyupdown}
\end{equation}
where $\D_\mu$ is the covariant derivative, ${T}^\mu_\nu$ is the energy--momentum tensor of
matter. The trace equation is  useful to get exact solutions:
\begin{equation}
M_P^2R- \sum_{n=1}^\infty
f_n\sum_{l=0}^{n-1}\left(\partial_\mu\Box^l  R  \partial^\mu\Box^{n-l-1}  R
+2\Box^l  R  \Box^{n-l}  R\right)-  6\Box\Fc(\Box)R=4\Lambda-{T}^\mu_\mu\, .
\label{eqEinsteinRonlytrace}
\end{equation}

\section{The Ansatz for Finding Exact Solutions}

It has been shown in~\cite{BMS,BKM,BKMV,KVSQS2011} that the following ansatz
\begin{equation}
\Box R=r_1R+r_2,
 \label{ansatzR}
\end{equation}
with constants $r_1\neq 0$ and $r_2$, is useful in finding exact solutions.

If the scalar curvature $R$ satisfies (\ref{ansatzR}), then equations (\ref{eqEinsteinRonlyupdown}) are
\begin{equation}
\begin{split}
&\frac12\left[M_P^2+2\left(\Fc(r_1)R+\frac{r_2}{r_1}(\Fc(r_1)-f_0)\right)\right]
\left(2R^\mu_\nu-\delta^\mu_\nu R\right)={T}^\mu_\nu+{}\\
+{}&\Fc'(r_1)\left[\partial^\mu R \partial_\nu
R-\frac{\delta^\mu_\nu}{2}\left(g^{\sigma\rho}\partial_\sigma R \partial_\rho
R+r_1\left(R+\frac{r_2}{r_1}\right)^2\right)\right]-\Lambda \delta^\mu_\nu
{}+\\+{}
&2\Fc(r_1)\left[\D^\mu\partial_\nu
R-\delta^\mu_\nu(r_1R+r_2)\right]-\frac{\delta^\mu_\nu}{2}\left[\Fc(r_1)R^2-\frac{r_2^2}{r_1^2}(\Fc(r_1)-f_0)
\right],
\end{split}
\label{eqEinsteinAnzatz}
\end{equation}
where $\Fc'$ is the first derivative  of $\Fc$ with respect to the argument.

We proceed to consider a traceless radiation along with a cosmological constant.
Under condition
(\ref{ansatzR}) the trace equation with ${T}^\mu_\mu=0$ becomes especially simple:
\begin{equation}
AR+\Fc'(r_1)\left(2r_1R^2+ \partial_\mu R \partial^\mu R\right)+B=0\,,
\label{Trequ_anzats}
\end{equation}
where constants $A$ and $B$ are defined as follows:
\begin{equation*}
A=4\Fc'(r_1)r_2-M_P^2-2\frac{r_2}{r_1}(\Fc(r_1)-f_0)+6\Fc(r_1)r_1,\quad
B=4\Lambda+\frac{r_2}{r_1}M_P^2+\frac{r_2}{r_1}A.
\end{equation*}
The simplest way to get a solution to equation (\ref{Trequ_anzats}) is
to impose
$\Fc'{(r_1)}=0$ and to put  $A=B=0$. These relations  fix values of $r_1$, $r_2$ and
the cosmological constant:
\begin{equation}
\label{r12}
r_2={}-\frac{r_1[M_P^2-6\Fc(r_1)r_1]}{2[\Fc(r_1)-f_0]},\qquad
\Lambda={}-\frac{r_2M_P^2}{4r_1}=M_P^2\frac{[M_P^2-6\Fc(r_1)r_1]}{
8[\Fc(r_1)-f_0]}.
\end{equation}

Under conditions $A=B=\Fc'(r_1)=0$ the complete set of equations~(\ref{eqEinsteinAnzatz}) simplifies to
\begin{equation}
\begin{split}
&2\Fc(r_1)(R+3r_1)
G^\mu_{\nu}={T}^\mu_\nu+
2\Fc(r_1)\left[g^{\mu\rho}\D_\rho\partial_\nu
R-\frac14\delta^\mu_\nu\left(
R^2+4r_1R+r_2\right)\right] \ .
\end{split}
\label{eqEinsteinAnzatz_Ai0}
\end{equation}
In general one is required to include radiative sources to get exact solutions of all equations~\cite{BKM}.
Let us emphasize that equations are general and we do not take into account the
properties of the metric.

\section{Relation between the anzats and  $R^2$ modified gravity}

It is well-known~\cite{R2_gravity} that the $f(R)$ gravity model, described by the action:
\begin{equation}
S_f=\int d^4x\sqrt{-g}\left(\frac
 {M_P^2}{2}f(R)+\mathcal{L}_\mathrm{M}\right),
 \label{f_R_action}
\end{equation}
has the following equations:
\begin{equation}
M_P^2\left(f^{(1)}(R)R^\mu_\nu-\frac{1}{2}f(R)\delta^\mu_\nu- (g^{\mu\rho}\D_\rho\partial_\nu-\delta^\mu_{\nu}
\Box)f^{(1)}(R)\right)=T^\mu_\nu,
\label{equ_f_R}
\end{equation}
where $f^{(1)}(R)$ is the first derivative of $f(R)$ with respect to $R$.

From (\ref{equ_f_R}) for a traceless matter, we get the following trace equation:
\begin{equation}
\label{FRtrace}
f^{(1)}(R)R-2f(R)+3\Box f^{(1)}(R)=0.
\end{equation}
One can see that at
\begin{equation}
\label{f_R}
f(R)=\frac{\Fc(r_1)}{M_P^2}\left[R^2+6r_1R+3r_2\right],
\end{equation}
equation (\ref{FRtrace}) is coincide with the ansatz (\ref{ansatzR}). Moreover, equations (\ref{eqEinsteinAnzatz_Ai0}) are equivalent to (\ref{equ_f_R}). Note that condition (\ref{r12}) gives the following connection:
\begin{equation*}
M_P^2=\frac{2}{r_1}\left[3\Fc(r_1)r_1^2-(\Fc(r_1)-f_0)r_2\right].
\end{equation*}

We proved that any solution of the modified gravity model (\ref{f_R_action}) with $f(R)$ given by (\ref{f_R}) and traceless matter is a solution of the initial system of equations~(\ref{eqEinsteinRonlyupdown}) on condition that $A=B=\Fc'(r_1)=0$ and the anzats (\ref{ansatzR}) is satisfied.

\section{Conclusion}
We have shown that any solution of the $R^2$ modified gravity model
(\ref{f_R_action}) either without matter, or with the radiation is a solution of
the corresponding SFT inspired non-local gravity models~(\ref{nlg_action}). We
do not assume some special form of the metric to prove this result. Note that
the initial nonlocal model is not totally equivalent to an $R^2$ model. Full
consideration of the model includes both the background
solution, and perturbations, which may not satisfy the ansatz (\ref{ansatzR})
in general.  The analysis of the cosmological perturbations in the considered
non-local models is given in~\cite{BKMV}. Note that, the $R^2$
modified gravity is very well studied~\cite{R2_gravity} and looks as a
realistic modification of gravity. In particular, the modern cosmological and
astrophysical data~\cite{PlanckInflation} confirm the predictions of the
Starobinsky inflationary model~\cite{Starobinsky}.

\medbreak

\noindent {\bf Acknowledgements.} This work is supported in part by the RFBR grant 14-01-00707. A.K. is supported in part by FWO-Vlaanderen through project G020714N, by the Belgian Federal Science Policy Office through the Interuniversity Attraction Pole P7/37, and by the Vrije Universiteit Brussel through the Strategic Research Program ``High-Energy Physics''.

\end{document}